\documentclass{aa}  

\usepackage{graphicx}
\usepackage{txfonts}
\usepackage{soul}

\usepackage[colorlinks=true, citecolor=blue]{hyperref}
\usepackage{subcaption}
\begin{document} 

   \title{Binary Stars as Dynamical Tracers in Globular Clusters .I. }\subtitle{First Observations of Bimodal Spatial Distributions}
   \author{M. Cadelano \inst{1,2} \thanks{Based on observations collected at the European Southern Observatory under ESO programmes 093.D-0228, 097.D-0145(A), 106.219K.001, 109.2305.001,  111.24LZ.001 (PI: Dalessandro) and at the Large Binocular Telescope under programme T-2021B-039 (PI: Cadelano).}\and
          E. Dalessandro \inst{2} \and
          J. Bruce \inst{3} \and
          E. Vesperini \inst{3} \and
          F. R. Ferraro  \inst{1,2} \and
          B. Lanzoni  \inst{1,2} \and   
          G. Beccari \inst{4} \and
          C. Giusti  \inst{1,2} \and
          F. Cusano \inst{2}  \and
          D. Paris \inst{5}
          } 

   \institute{Department of Physics and Astronomy ‘Augusto Righi’, University of Bologna, via Gobetti 93/2, I-40129 Bologna, Italy
    \and
    INAF – Astrophysics and Space Science Observatory of Bologna, via Gobetti 93/3, I-40129 Bologna, Italy
    \and
             Department of Astronomy, Indiana University,Swain West, 727 E. 3rd Street, IN 47405 Bloomington, USA
    \and    
            European Southern Observatory, Karl-Schwarzschild-Strasse 2, 85748 Garching bei München, Germany
    \and    
            INAF – Osservatorio Astronomico di Roma, via di Frascati 33, 00078 Monte Porzio Catone, Italy
}

   \date{}

 
\abstract{

{We present the first homogeneous study of the radial distribution of the binary fraction across the full extent of six Galactic globular clusters (GCs) spanning a wide range of dynamical ages, from dynamically young systems to core-collapsed clusters. We measured the radial variation of the binary fraction using a combination of deep optical HST observations and wide-field ground-based data. For the first time, we provide evidence that the binary fraction in GCs does not decrease monotonically with radius, as commonly assumed, but instead exhibits a bimodal distribution characterized by an excess in the outer regions. Specifically, the binary fraction displays a central peak, followed by a minimum at intermediate radii and a rising branch beyond approximately 1–2 half-light radii. The position of this minimum correlates with the cluster relaxation timescale, indicating that it is shaped by long-term dynamical effects of two-body relaxation driving the binary evolution, segregation, and disruption. 
The minimum radius also correlates with the $A^+_{rh}$ parameter derived from the radial distribution of blue straggler stars, an empirical indicator of the cluster dynamical age, further supporting the interpretation that this feature is due to internal dynamical processes. Numerical simulations presented in a companion paper show that such bimodal distributions naturally arise from the combined effects of binary disruption and mass segregation of the surviving binaries in clusters.
}}


   \maketitle

\section{Introduction} \label{sec:intro}

{Binary stars play an important role in shaping the long-term evolution of dense stellar systems.} 
In fact, in collisional environments such as globular clusters (GCs), where stellar gravitational interactions are frequent, binaries play a crucial role in shaping both the dynamical evolution of the system and the properties of its stellar populations. In particular, in the inner and denser regions of a cluster, interactions between stars and binaries progressively alter the binary population: new binaries can form, while existing ones may be disrupted or modified \citep[e.g.][]{hut92,hurley07,fregeau09}. This process is responsible for the formation of a variety of exotic objects, such as blue straggler stars, cataclysmic variables, X-ray binaries, and millisecond pulsars \citep[e.g.][]{ferraro09,ferraro23nat,Riverasandoval18,cadelano17_ngc6440,cadelano18,cadelano22,Zhao22,douglas22,zhang23,ettorre25,dutta25,lian25}, whose frequency is enhanced in GCs with respect to the field population. {Among these systems, blue straggler stars, primarily formed through main-sequence binary evolution \citep[e.g.,][]{reggiani25,ferraro26}, have proven to be effective probes of cluster internal dynamics \citep{ferraro12,ferraro18,Ferraro19,ferraro23,billi23,billi24,billi26,giusti24,giusti25}.}
{Binary stars also play an important role in  the dynamics of the  multiple stellar populations found in most GCs (see e.g. \citealt{gratton19} for a review). Numerical studies \citep{vesperini11, hong15, hong16, hong19, Hypki22, sollima22,bruce26}, have predicted that the initial differences between the spatial distributions of the first-population stars (FP; those with chemical properties similar to field stars) and the second-population stars (SP; those with anomalous abundances in various light elements such as Na, O, Al, Mg, C, N, and He) can lead to significant differences between the survival rate of FP and SP binaries and in the evolution and present-day properties of the surviving FP and SP binary population. 
A few  observational studies have started to investigate the properties of FP and SP binaries and revealed differences  generally consistent with the theoretical expectations \citep{lucatello15, dalessandro18, kamann2020, milone20, milone25, bortolan25}.}

In general, gravitational interactions are expected to cause binaries to sink toward the cluster center due to the effect of dynamical friction (hence, mass segregation), as they are, on average, more massive than single stars. They also make hard binaries harder, providing an energy source that can potentially slow down and even prevent the core collapse \citep{Goodman89, mcmillan11, vesperini94, heggie06, trenti07}.{Modern dynamical studies suggest that binaries containing stellar-mass black holes may play a particularly important role in this process in many GCs \citep[e.g.,][]{heggie13,heggie14,kremer18,askar18}.} The typical timescale of these dynamical processes (i.e., the relaxation time) depends on the cluster’s physical properties and on the distance from the cluster center and in the outskirts it can be comparable to or even exceed a Hubble time. This suggests that the binary distribution can vary locally within a cluster, and that cluster-to-cluster differences in binary properties are expected due to variations in structural parameters. 
On the theoretical side, several studies have investigated the evolution of the binary population in GCs using different approaches. \citet{sollima08} analyzed the evolution of the binary content through a fully analytical approach, finding that ionization and evaporation are the main processes driving the evolution of the binary fraction. 
Both Monte Carlo simulations \citep[][see also \citealt{ivanova05}]{fregeau09} and N-body simulations \citep{hurley07,trenti07,geller13b,geller13a,Hypki22} found that while the hard binary fraction in the core generally increases, the overall binary fraction evolves in a way that depends on the balance between the rate of escape of single stars and the loss of binary stars due to disruption and escape.
\citet{geller13b} showed that the combined effect of soft-binary disruptions and mass segregation can produce, after roughly one relaxation time, a bimodal binary distribution (see their Figure 2) although the radial variation in the binary fraction associated to the bimodality found in their study is very mild. 
This profile is characterized by a central peak, generated by the most segregated binaries, followed by a decreasing portion reaching a minimum in the cluster's intermediate regions and a rising branch in the outskirts. 
As time progresses, the minimum drifts outward, but the bimodality is rapidly erased. {In fact, after approximately
three relaxation times the distribution becomes monotonically
decreasing for the rest of the cluster's evolution. In the second paper of this work (Bruce et al. 2026b - hereafter Paper II) we further explored the effects of the clusters internal dynamics, binary disruption and segregation on the evolution of the radial profile of the binary star fraction considering the more complex structural properties of clusters hosting multiple stellar populations. The results of our analysis reveals the development of a significant bimodality generally consistent with the observational findings presented in this paper (see Paper II for further details). }

From the observational side, however, despite their importance as a tool to understand both the cluster internal dynamics and the impact of the latter in the cluster stellar and binary populations, the properties of binaries, particularly their radial distribution, remain poorly constrained due to the demanding observational requirements. Three main techniques are commonly used to estimate the binary fraction in star clusters. The first relies on detecting binaries through radial velocity variability \citep[e.g.,][]{lucatello15,dalessandro18,Torres21,Geller21,saracino23,wragg24,mullerhorn25}; the second involves identifying eclipsing binaries \citep[e.g.,][]{prsa11,kirk16,rozyczka22}; and the third exploits the distribution of stars along the main sequence in color-magnitude diagrams (CMDs) \citep[e.g.,][]{bellazzini02,sollima07,milone12,dalessandro11,dalessandro15}. However, the first two methods are strongly biased toward systems with high inclination angles and short orbital periods, leading to a significant underestimation of the binary fraction (see, for example, the discussion in \citealt{lucatello15,dalessandro18}). The third approach, on the other hand, takes advantage of the fact that the flux of unresolved binaries equals the sum of the fluxes of their two components. As a result, binaries composed of two main sequence stars appear shifted toward brighter magnitudes compared to single main sequence stars. While this method is limited to main sequence binaries, it provides large statistical samples and is free from biases related to binary orbital periods and inclination angles. {On the other hand, this method is more sensitive to binaries with relatively large mass ratios, while systems with low-mass companions remain indistinguishable from single stars. As a consequence, estimates of the total binary fraction generally require an extrapolation below a minimum detectable mass ratio.} This technique, applied to a large sample of GCs \citep{sollima07,milone12} revealed that binaries are centrally concentrated, as expected from mass-segregation processes, and that the blue straggler fraction correlates with the binary fraction. However, these studies are limited to the cluster inner regions sampled by the Hubble Space Telescope (HST).
The radial distribution of binaries across a significant portion of a Galactic GC’s extent has been studied in only few cases so far\footnote{see \citealt{Mohandasan24} for the case of Magellanic Cloud star clusters}: NGC~288 \citep{bellazzini02}, NGC~6254 \citep{dalessandro11}, NGC 5466 \citep{beccari13}, and NGC 6101 \citep{dalessandro15}, with strikingly different results. In NGC~288 and NGC~6254, the binary fraction is sampled within 2 half-mass radii and it decreases monotonically from $\sim0.1-0.2$ in the core to $\sim0.015-0$ at a distance between 1 and 2 half-mass radii. {NGC 5466 hosts a binary fraction of  $\sim0.06-0.07$ within the half-light radius, decreasing to $\sim0.055$ at larger radii, with a hint of a bimodal distribution similar to that observed for blue straggler stars \citep{beccari13}. In particular, both distributions display a minimum around one half-light radius.} Finally, NGC~6101 displays a constant binary fraction of $\sim0.14$ across the entire cluster. A similar trend is found for blue straggler stars, suggesting
that mass segregation in NGC~6101 has not yet significantly influenced the observed binary population and blue straggler stars.
The differences in the observed binary radial distributions among these clusters likely result from varying degrees of dynamical evolution due to potentially different initial conditions. However, the very limited sample available so far prevents a comprehensive characterization of the binary population and its evolution within GCs. The aim of this study is to adopt a homogeneous approach to investigate the radial variation of the binary fraction across the whole extension of a representative sample of old GCs spanning a broad range of properties, including mass, density, and relaxation time. The selected clusters are NGC~104 (47 Tucanae), NGC~5053, NGC~6218 (M~12), NGC~6981 (M~72), NGC~7078 (M~15), and NGC~7099 (M~30). Table\ref{tab:gc} summarizes the key properties of these targets.
{The clusters were selected based on the availability of deep HST photometry for the central regions, ultra-deep wide-field observations sampling the external regions well beyond the cluster tidal radii, and the corresponding artificial-star catalogues required for completeness corrections. The final sample spans a broad range of masses, densities, structural parameters, and dynamical ages representative of Galactic globular clusters.}

The paper is structured as follows: Section~\ref{sec:data} describes the dataset and data reduction procedure; Section~\ref{sec:fbin} outlines the techniques used to measure the binary fraction. The results are discussed in Section~\ref{sec:results}, while Section~\ref{sec:conc} provides a summary and final conclusions.

\section{Dataset}
\label{sec:data}

\begin{table*}
\caption{Main properties of the clusters analyzed in this work. From top to bottom: the cluster gravitational center, age, core density, mass, distance from the Sun, core ($r_c$), half-light ($r_h$) and tidal radii ($r_t$), concentration, central and half-mass relaxation timescales, dynamical ages (number of relaxations experienced by the cluster in units of central and half-mass relaxations), slope of the global mass-functions. \label{tab:gc}}
\centering
\begin{tabular}{lcccccc}
\hline\hline
Parameter & NGC 104 & NGC 5053 & NGC 6218 & NGC 6981 & NGC 7078 & NGC 7099 \\
\hline
RA (J2000)  & $00^h24^m05.71^s$ &  $13^h16^m26.70^s$ & $16^h47^m14.24^s$ &  $20^h53^m27.71^s$ & $21^h29^m58.31^s$ & $21^h40^m22.13^s$ \\
Dec (J2000)  & $-72^{\circ}04\arcmin52.20\arcsec$ & $17^{\circ}42\arcmin0.59\arcsec$ & $-1^{\circ}56\arcmin53.90\arcsec$ &  $-12^{\circ}32\arcmin14.3\arcsec$ & $12^{\circ}10\arcmin1.4\arcsec$ & $-23^{\circ}10\arcmin47.4\arcsec$ \\
$t_{age}$ (Gyr) & $12.75\pm0.50$ & $13.50\pm0.75$ & $13.25\pm0.75$ & $12.75\pm0.75$ & $13.25\pm0.75$ & $13.25\pm0.75$  \\
$\log \rho_c \ (M_{\odot} \ pc^{-3})$ & 4.72 & 0.50 & 3.13 & 2.3 & 7.12 & 5.91   \\
Mass ($10^5 \ M_{\odot}$) & $8.53\pm0.05$ & $0.74\pm0.16$ & $1.07\pm0.03$ & $0.69\pm0.12$ & $6.33\pm0.07$ & $1.43\pm0.06$  \\
$R_{SUN} \ (kpc)$ & $4.52\pm0.03$ & $17.5\pm0.2$ & $5.11\pm0.05$ & $16.66\pm0.18$ & $10.71\pm0.10$ & $8.46\pm0.09$ \\
$r_c \, (\arcsec)$ & $21.6$ & $124.8$ & $47.4$ & $27.6$ & $8.4$ & $3.6$  \\
$r_{h} \, (\arcsec)$ & $190.2$ &  $156.6$ & $106.2$ & $55.8$ & $60.0$ & $61.8$ \\
$r_t \, (\arcsec)$ & $2538$ &  $684$ & $1038$ & $450$ & $1638$ & $1140$ \\
$c$ & $2.07$ & $0.74$ & $1.34$ & $1.21$ & $2.29$ & $2.50$    \\
$\log t_{rc}$ (yr) & 7.84 & 9.81 & 8.19 & 8.72 & 7.84 & 6.37 \\
$\log t_{rh}$ (yr) & 9.55 & 9.87 & 8.87 & 9.23 & 9.32 & 8.88  \\
$t_{age}/t_{rc}$ & 184 & 2 & 86 & 24 & 192 & 5650 \\
$t_{age}/t_{rh}$ & 3.6 & 1.8 & 17.9 & 7.5 & 6.3 & 17.5 \\
$\alpha_{GLOBAL}$ & -0.45  & -1.26 & -0.36 & -0.7 & -1.07  & -0.68 \\
\hline
\hline
\end{tabular}
\tablefoot{Ages are from \citet{dotter2010}. Gravitational centers, central densities, masses and distances from the Sun are from \citet{baumgardt2018}. Cluster structural parameters are extracted or derived from \citet{harris2010}. The slopes of the global mass functions are from \citet{webb17,Ebrahimi20,cadelano20}.}
\end{table*}

\begin{figure*}[h] 
\centering
\includegraphics[scale=0.29]{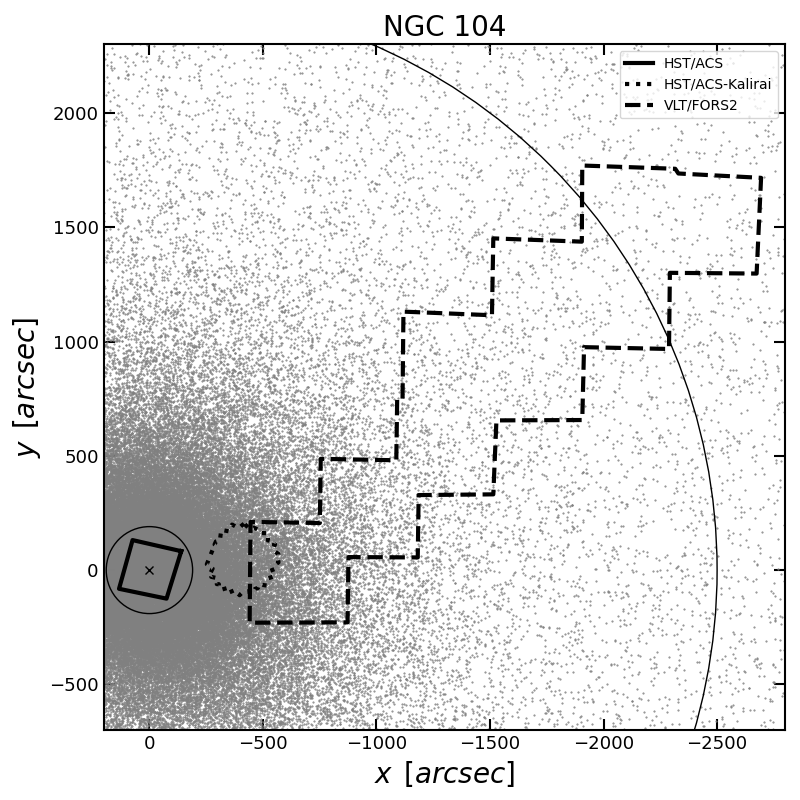}
\includegraphics[scale=0.29]{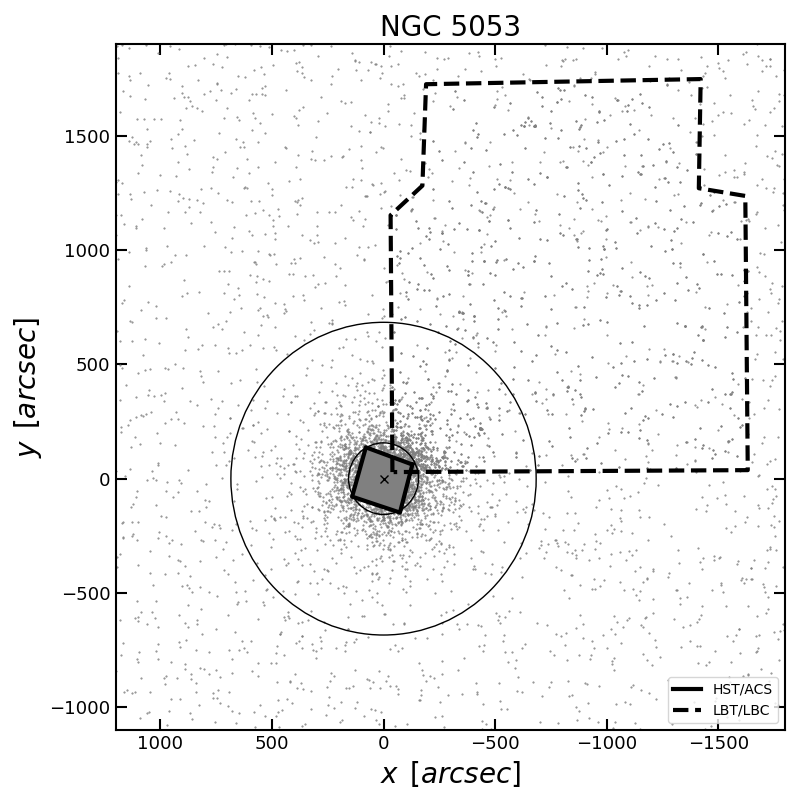}
\includegraphics[scale=0.29]{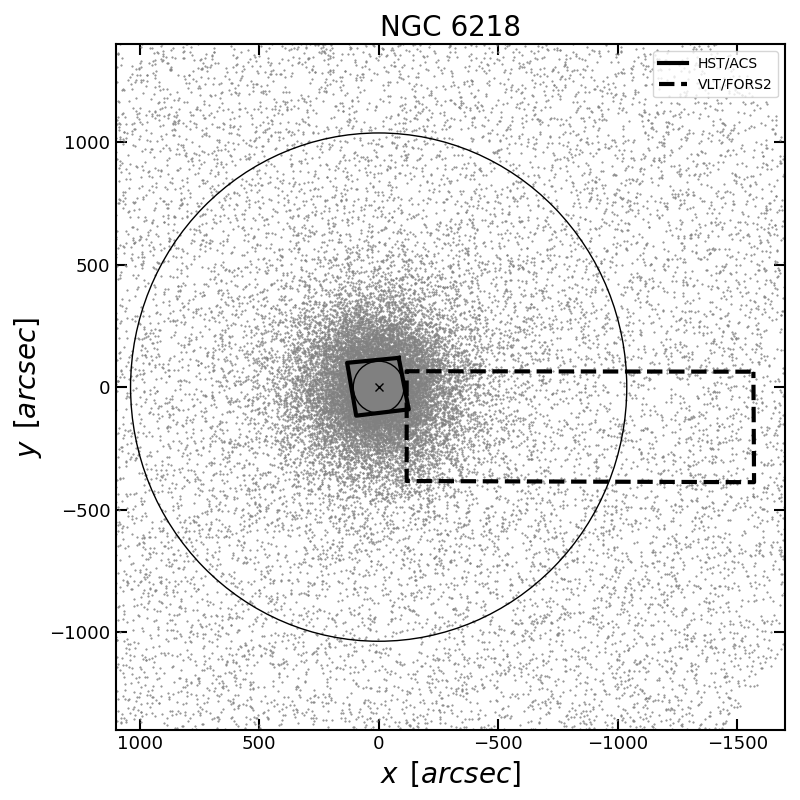}
\includegraphics[scale=0.29]{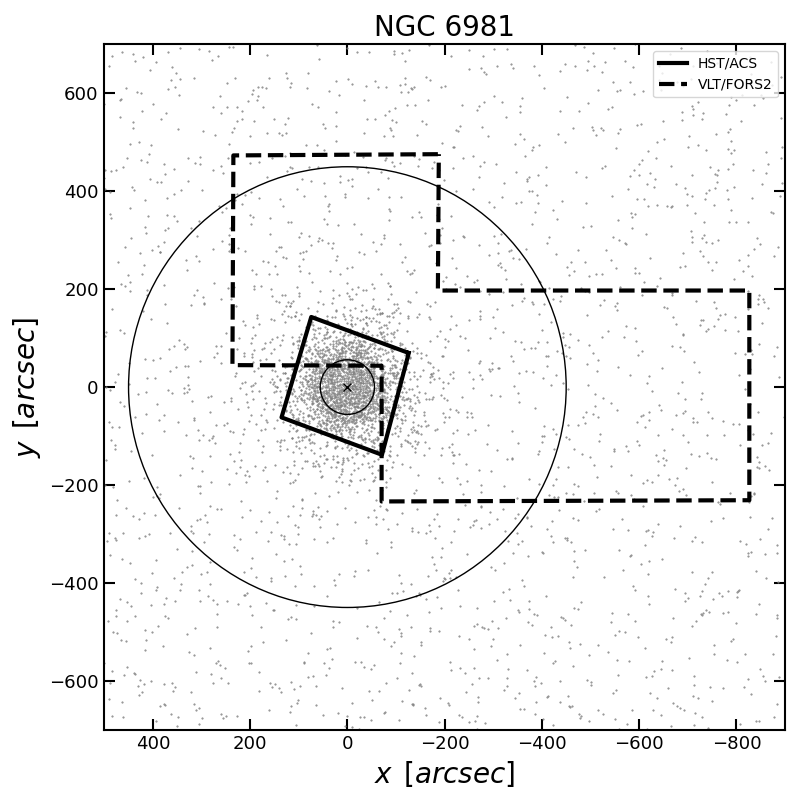}
\includegraphics[scale=0.29]{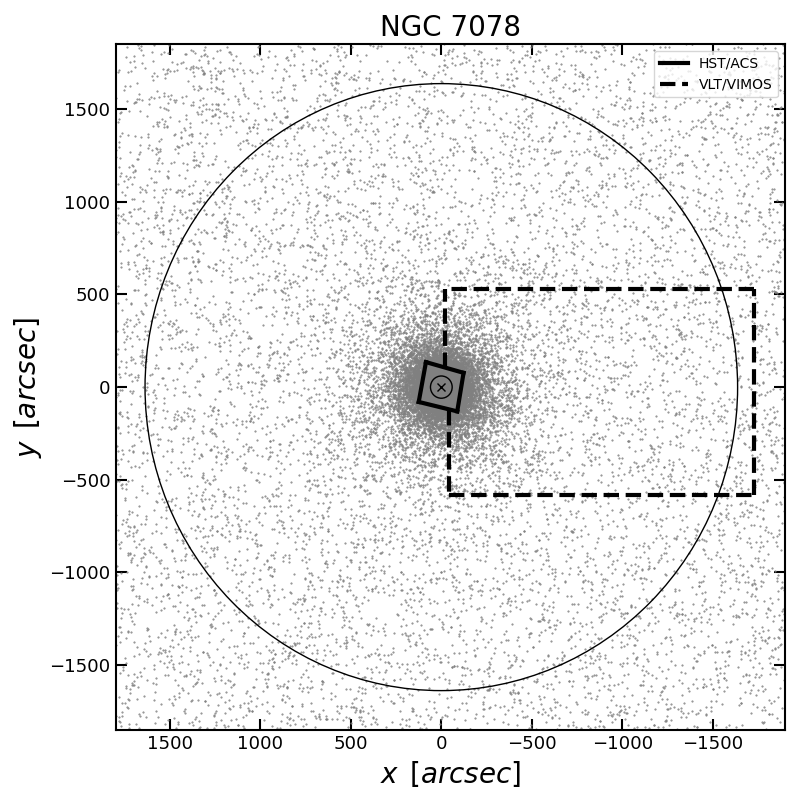}
\includegraphics[scale=0.29]{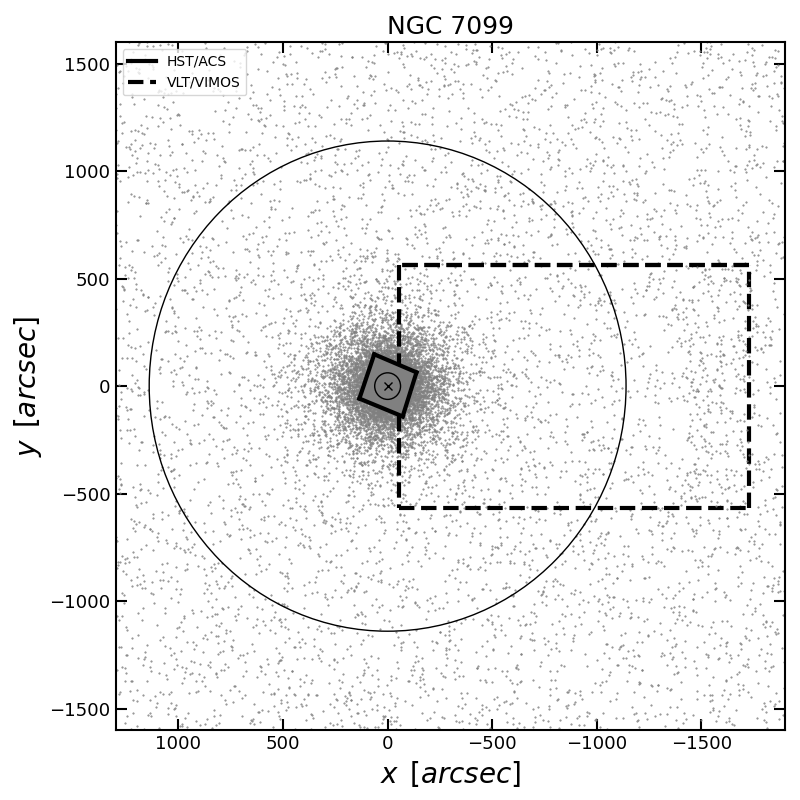}
\caption{Fields of view sampled by the different datasets adopted in this work. Gray points are clusters stars drawn from the Gaia DR3 catalogues \citep{GaiaDR3}.}
\label{fig:fov}
\end{figure*}

\begin{figure*}[h] 
\centering
\includegraphics[scale=0.29]{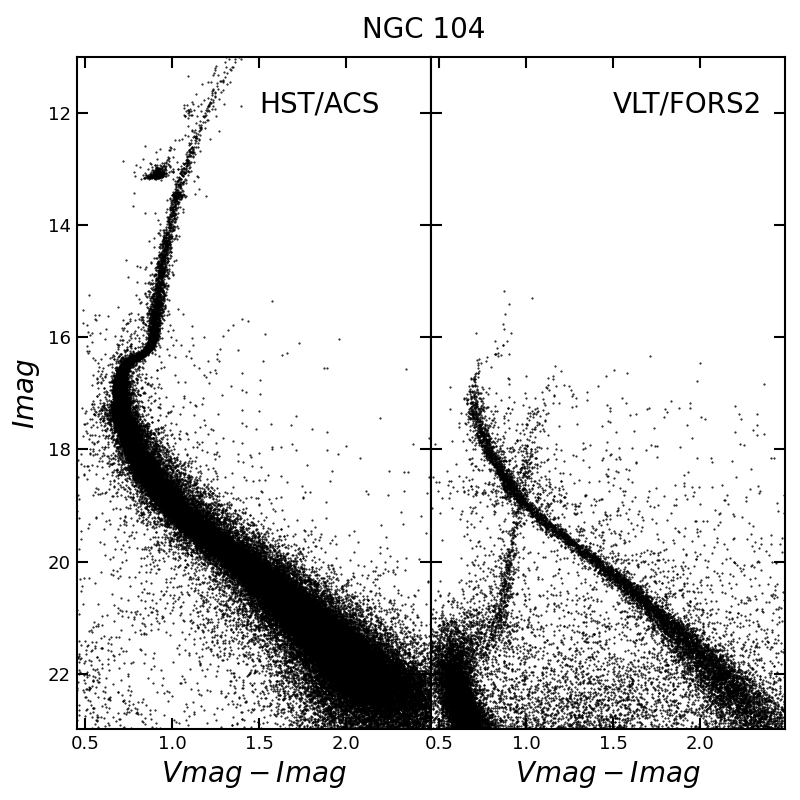}
\includegraphics[scale=0.29]{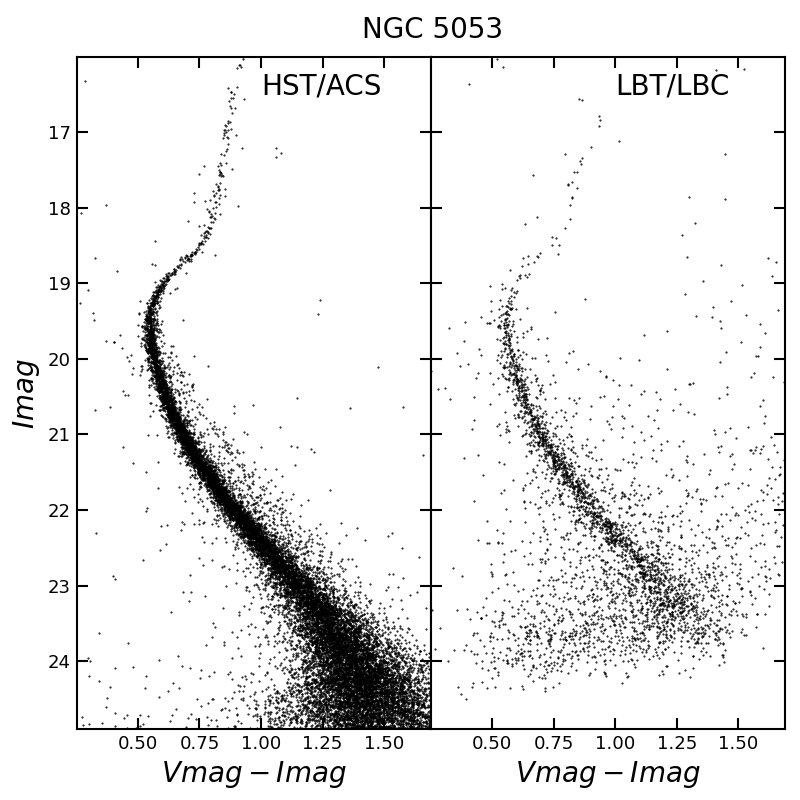}
\includegraphics[scale=0.29]{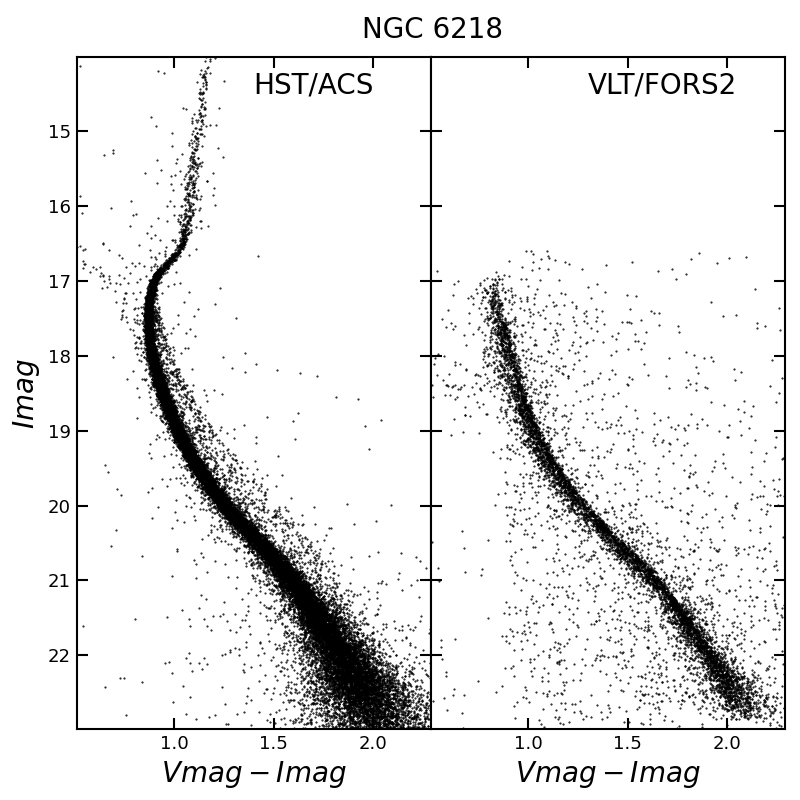}
\includegraphics[scale=0.29]{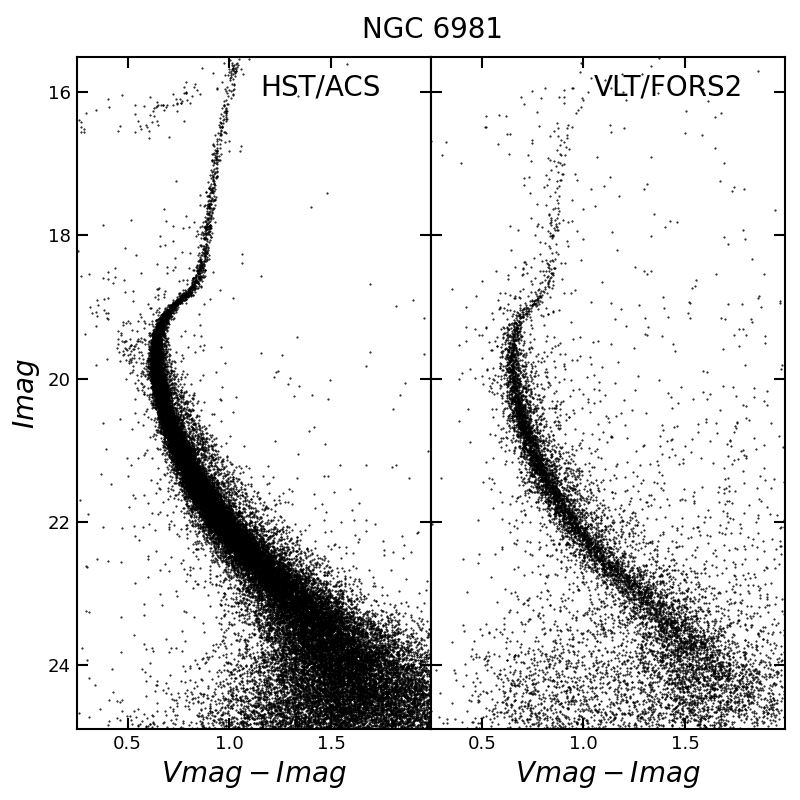}
\includegraphics[scale=0.29]{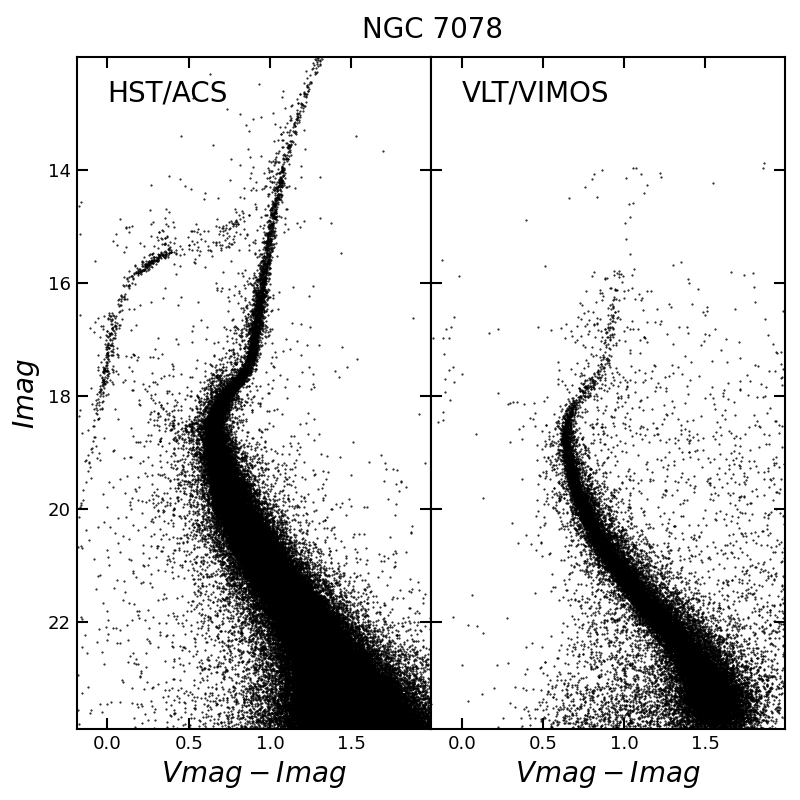}
\includegraphics[scale=0.29]{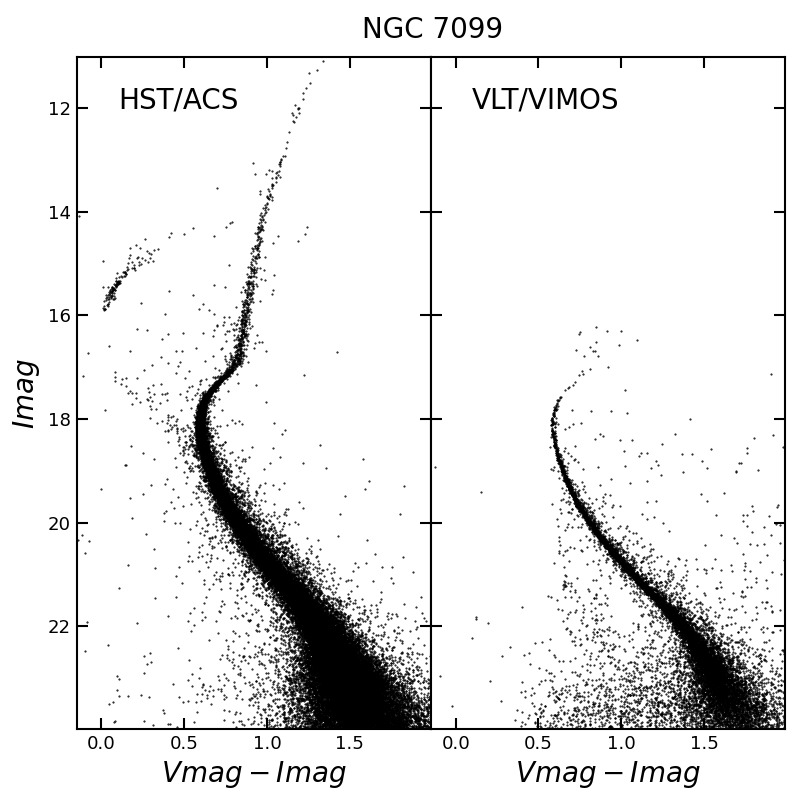}
\caption{CMDs of the datasets adopted in this work. For each cluster, the left-hand panel shows the CMD obtained through the ACS dataset, while the right-hand panel that obtained from the extremely deep FORS2/VIMOS/LBC dataset.}
\label{fig:cmd}
\end{figure*}

The work is based on a combination of different datasets for each cluster. The central regions are sampled using publicly available high-resolution observation catalogs obtained with the HST as a part of the {\it ACS survey of Galactic Globular Clusters} \citep{sarajedini07,anderson08}. The datasets are composed of images equally split between the F606W and F814W bands. The catalogues also provide calibrated Johnson V-band and I-band magnitudes, which we adopted throughout the whole work for homogeneity purposes with the other catalogues.

The cluster external regions are sampled by multi-instrument ultra-deep observations sampling the cluster main sequences down to $V\sim26$ across and beyond the cluster tidal radii listed in Table~\ref{tab:gc}. In particular, in the case of NGC~6218 and NGC~6981 we used V-band and I-band photometry obtained through VLT/FORS2 images  (Prop ID: 093.D-0228, PI:Dalessandro) and originally presented by \citet{sollima17}. Similarly, in the case of NGC~104, we used V-band and I-band images obtained with the VLT/FORS2 (Prop ID:  106.219K.001, 109.2305.001, PI:Dalessandro). The details of the dataset and of data reduction of this cluster will be presented in a forthcoming dedicated paper (Dalessandro et al. 2026, in preparation). Moreover, to fill the radial gap between the inner ACS and outer FORS2 dataset, we also used the publicy available HST catalogue presented in \citet{kalirai12} and composed of ultra-deep ACS observations in the F606W and F814W sampling the cluster main-sequence down to $\sim10$ mag below the turn-off. We refer the reader to the aforementioned paper for the dataset details and photometry. In the case of NGC~7078 and NGC~7099 we used V-band and I-band photometry obtained through the VLT/VIMOS camera (Program ID: 097.D-0145(A), PI: Dalessandro) and originally presented by \citet{cadelano20}. Finally, the external regions of NGC~5053 have been sampled using LBT/LBC observations (Prop ID: IT-2021B-039, PI: Cadelano) and presented here for the first time. {The NGC~5053 data set is composed of 50 V-band images with an exposure time of 205 s, and 25 I-band images with an exposure time of 180 s, acquired with seeing in the range $0.8\text{–}2.2\arcsec$ and at an airmass of $\sim 1.1$. All images were pre-processed using the dedicated LBC pipeline.} The photometry was performed using DAOPHOT II \citep{stetson87} following the same procedure fully described in \citet{cadelano17,cadelano23,deras23}. {The quality of the LBT/LBC photometry was assessed by comparing the photometric uncertainties as a function of magnitude with those of the other ground-based datasets adopted in this work. We verified that the resulting photometric errors are fully comparable and therefore suitable for a homogeneous  analysis across the whole cluster sample.} The instrumental positions were transformed to the absolute system by using the stars in common with the Gaia Data Release 3 \citep{GaiaDR3}. The instrumental magnitudes were reported to the Johnson photometric system by using the stars in common with the wide-field catalog described by \citet{stetson19}. The field of views covered by the observations used in this work and the CMDs are presented in Figure~\ref{fig:fov} and Figure~\ref{fig:cmd}, respectively. {In the regions where different datasets overlap, the photometric calibrations were verified using stars in common. No significant systematic offsets were detected, and duplicate sources were removed before constructing the final catalogues used in the analysis.}

To study the cluster binary fractions and their radial variation it is necessary to take into account the completeness level of each adopted catalog for stars with different magnitudes and located at different distances from the cluster centers. {For all  clusters except NGC~5053, we used the completeness values derived in \citet[][Dalessandro et al. 2026, in prep.]{sollima17,cadelano20}, which make use of proprietary artificial star catalogs for the ground-based observations and publicly available ones for the ACS observations \citep{anderson08,kalirai12}}.
{Finally, in the case of NGC~5053, for the ACS dataset we used the artificial star catalogs provided along with the main catalogs of the ACS Survey of Galactic Globular Clusters (see Section 6 of \citealt{anderson08}). In the case of the LBC dataset,} we performed a large number of artificial star experiments following the same prescriptions described in detail in \citet[][see also \citealp{dalessandro15,bellazzini02}]{cadelano20}. {Briefly, we created a list of artificial stars with input magnitudes extracted from the observed I-band luminosity function, extrapolated beyond the limiting magnitude, and assigned the corresponding V-band magnitudes through the observed mean ridge line.} These
artificial stars were added to the real images by using the
DAOPHOT/ADDSTAR software. The photometric reduction process and is exactly the same as described before. The process was iterated multiple times at the end of the runs, about 100000 were simulated along the entire LBT/LBC
field of view. 

We assigned to each star in each catalogue a completeness value $C=N_o/N_i$ , defined as the ratio between the number of stars recovered at the end of the artificial star test ($N_o$) and that of stars actually simulated ($N_i$). To account for the effect of crowding, the completeness $C$ of each star was derived by using only objects located within a radial bin centered on the location of the star and with a width of $5\arcsec$ for the ACS catalogs and $50\arcsec$ for the wide-field VIMOS, FORS2 and LBC catalogs. The bin widths were chosen as a compromise between having enough statistics and sampling a limited radial extension. Since the completeness level strongly depends also on the stellar magnitude, we evaluated $C$ considering only simulated objects within a 0.5 large magnitude bin, centered on the I-band magnitude of each star. 
{For reference, the 50\% completeness level is reached at approximately $\rm 20.5 < Imag < 24$ in the ACS datasets and at approximately $\rm 22 < Imag < 24$ in the wide-field datasets, depending on the cluster and radial distance from the cluster centre. In all cases, the binary-fraction analysis was restricted to magnitude ranges where the completeness exceeds 50\%.}

\section{The binary fraction} \label{sec:fbin}

As mentioned in Section~\ref{sec:intro}, a binary system in a GC appears as an unresolved source with a total flux equal to the sum of its components’ fluxes. This results in a systematic shift in magnitude, depending on the relative brightness of the two stars. {We show in Figure~\ref{fig:fminbox} the CMD of one of the clusters studied in this work. The dashed line shows the cluster mean ridge line, calculated as the average stellar color as a function of magnitude. This line represents the ideal location of single main-sequence stars. Any deviation from the mean ridge line is either due to photometric uncertainties and/or blending from unresolved binaries\footnote{The effect of multiple populations is expected to be negligible in the adopted filter combination, while it becomes important when using specific near-UV or infrared filter combinations \citep[e.g.,][]{cadelano23}.}}. Binary systems composed of two main-sequence stars form a “secondary main sequence” on the red side of the single-star main sequence. The magnitude shift is related to the mass ratio $q = M_2 / M_1$: systems with $q = 1$ produce a maximum shift of $\sim0.75$ mag, while smaller $q$ values lead to shifts that may be eventually indistinguishable from single main sequence stars due to photometric errors. Hence, only binaries with $q$ larger than a given threshold $q > q_{min}$ can be photometrically identified. In this work, we estimate both a lower limit on the binary fraction, counting systems with $q > q_{min}$, and the total binary fraction by modelling the mass-ratio distribution $f(q)$ and comparing observed and synthetic CMDs.

To properly quantify the binary fraction, it is essential to account for contamination from blended sources and field interlopers. Blending arises from the chance superposition of two stars, mimicking the brightness enhancement seen in binaries. We estimate this contamination statistically using artificial star tests, measuring asymmetries in the residuals between input and output magnitudes. Field interlopers contaminate both the single and secondary MS, and their contribution is quantified using observations beyond the cluster tidal radius.

\subsection{Minimum binary fraction}
\label{fmin}
\begin{figure}[h] 
\centering
\includegraphics[scale=0.3]{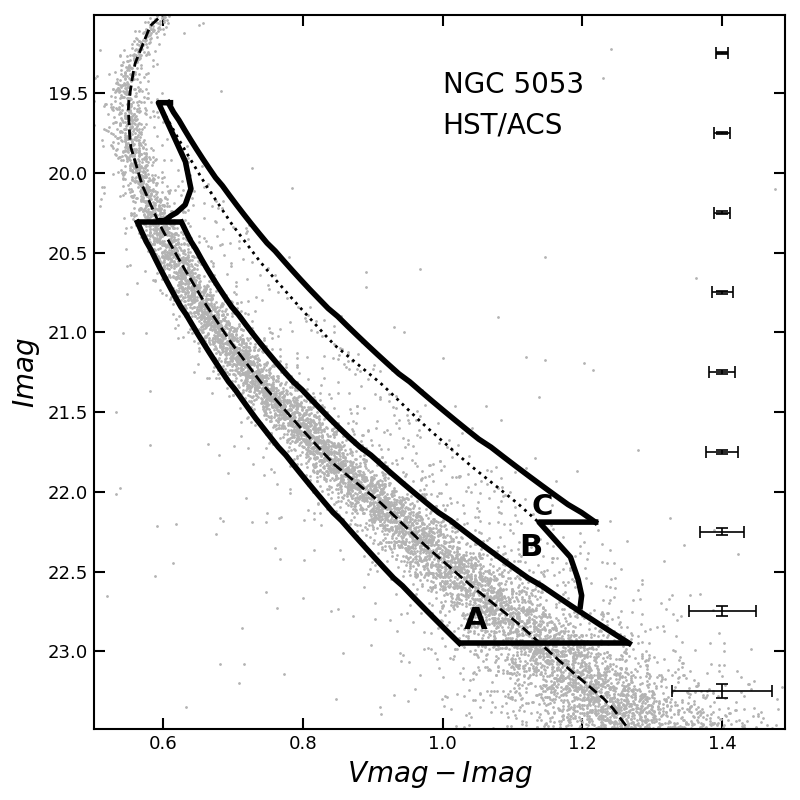}
\caption{CMD of NGC~5053 with an example of the selection boxes used to select the MS stars (box A) and binary stars (boxes B+C). The dashed line marks is the MS mean ridge line, while the dotted line marks the equal-mass binary sequence (see Section~\ref{fmin}). The black errorbars show the average photometric errors at different magnitudes.}
\label{fig:fminbox}
\end{figure}

{We define the binary fraction $\xi$ as the ratio between the number of binary systems and the total number of stellar systems (single stars plus binary systems) in a given magnitude (mass) range:}

\begin{equation}
\label{eq:fbin1}
\xi = \frac{N_b}{N_{MS} + N_b}
\end{equation}

where $N_b$ and $N_{MS}$ are the numbers of binaries and single main sequence stars, respectively. 
We select stars within an I-band magnitude ranging from 0.5 to 3 mag below the turn-off, where completeness exceeds 50\% at any clustercentric distance and approximately corresponding to stars with masses in the range $0.5M_{\odot}-0.75 M_{\odot}$. We consider a star to be a genuine single object if its measured I-band magnitude lies within this range and its distance in color from the mean ridge line is less than three times the photometric uncertainty (box $A$ in Figure~\ref{fig:fminbox}). To identify binaries, we convert the magnitude range into a mass range using appropriate isochrones from the Dartmouth Stellar Evolution Database \citep{dotter08}. The CMD positions of binaries with different $q$ values are then mapped, forming a sequence extending up to a maximum 0.75 mag shift from the mean ridge line (box $B$ in Figure~\ref{fig:fminbox}). We account for photometric errors by extending this region (box $C$). The amount of extension is calculated by measuring the average color errors as a function of magnitude.  In the case of the investigated clusters, the color spread induced by photometric errors along the main sequence makes possible the direct detection of binaries with a minimum mass-ratio $q_{min} \approx 0.4-0.5$, while those with lower mass rations are indistinguishable from single main sequence stars. This lower limit on the binary fraction is hereafter referred to as the minimum binary fraction, $\xi_{min}$. {The binary-selection boxes are defined once for each cluster and are kept fixed throughout the entire analysis. Consequently, the minimum detectable mass ratio $q_{min}$ remains constant across all radial bins of a given cluster.}

We divide the field of view into concentric annuli and in each of them we define $N_{MS}$ as the completeness-corrected number of stars in box $A$, and $N_b$ as the completeness corrected number of binaries in boxes $B + C$. We correct for blended-source contamination ($N_{blend}$) by counting the number of stars inside box $B+C$ in the artificial star CMDs, normalized by the ratio of real to artificial stars in box $A$. Field interloper contamination ($N_b^{field}$ and $N_{MS}^{field}$) is estimated from the density of stars beyond the tidal radius inside box $B+C$ and $A$, respectively, multiplied by the bin area.

Therefore, we refine Equation~\ref{eq:fbin1} as:

\begin{equation}
\label{eq:fbin2}
\xi_{min} = \frac{N_b - N_b^{field} - N_{blend}}{(N_{MS} - N_{MS}^{field}) + (N_b - N_b^{field} - N_{blend})}
\end{equation}

Uncertainties on $\xi_{min}$ are evaluated through error propagation, assuming Poisson statistics.

\subsection{Total binary fraction}
\label{sec:fbintot}

To determine the total binary fraction ($\xi_{TOT}$), we compared the observed CMD with a set of synthetic CMDs generated for different input binary fractions ($\xi_{IN}$).

For each $\xi_{IN}$ in a given radial annulus, we drew a sample of $N$ stars from the artificial star catalog, replicating the observed luminosity functions, thus naturally accounting for blending and photometric incompleteness. From this sample, $N_b = N\xi_{IN}$ stars were treated as binary systems. We assigned primary masses by randomly sampling the cluster's global mass function \citep{webb17,Ebrahimi20,cadelano20} and derived the secondary masses by drawing mass ratios from the distribution of \citet{fisher05}. {The global mass functions adopted for each cluster are listed in Table~\ref{tab:gc}. We verified that the use of local, radially varying mass functions instead of the global one has a negligible impact on the inferred binary fractions, even in clusters showing strong mass segregation. The adoption of different mass-ratio distributions has a negligible impact on the inferred radial profile of the binary fraction. In particular, we verified that adopting a flat mass-ratio distribution does not significantly affect either the shape of the radial profile or the position of the minimum \citep[see also][]{sollima07,dalessandro11}.} The expected magnitudes of the binaries were then calculated using the mass-luminosity relation of the appropriate \citet{dotter08} isochrone. {To simulate field contamination, we {randomly} extracted $N_f = \rho_f A$ stars from the region beyond the tidal radius, where $\rho_f$ is the stellar density and $A$ is the annulus area. Therefore, the color-magnitude distribution of the contaminants is directly inherited from the observed field population. The magnitudes of the extracted stars are then perturbed according to the photometric error distribution appropriate for the radial bin being analysed, thus reproducing the observational scatter expected at that clustercentric distance.} We then compared the observed and synthetic CMDs by counting the ratio of single to binary stars within the selection boxes defined earlier (see Figure~\ref{fig:fminbox}). The whole process was repeated $N_{iter} \sim 100$ times to address stochastic variations. The typical outcome of this procedure is presented for the case of NGC~5053 in Figure~\ref{fig:ftotexample}. 

{We explored a grid of $\xi_{IN}$ values and, for each of them, generated a large number of Monte Carlo realizations. For every realization, we measured the quantity $r_{sim}=N_{B+C}/N_A$, defined as the ratio between the number of stars falling within the binary-star and single-star selection boxes. We then compared this distribution with the value measured in the observed CMD ($r_{obs}$). {The resulting distribution of residuals was described by a Gaussian function and combined with the probability distribution associated with the observational uncertainty, yielding a probability distribution as a function of $\xi_{\rm IN}$. Since this probability distribution is evaluated on a discrete grid of input binary fractions and is affected by the finite number of Monte Carlo realizations, it was finally fitted with a Gaussian function to estimate its central value and width. The mean and dispersion of the best-fitting Gaussian were adopted as the best-fit value of $\xi_{\rm TOT}$ and its uncertainty, respectively.}


\begin{figure}[h] 
\centering
\includegraphics[scale=0.32]{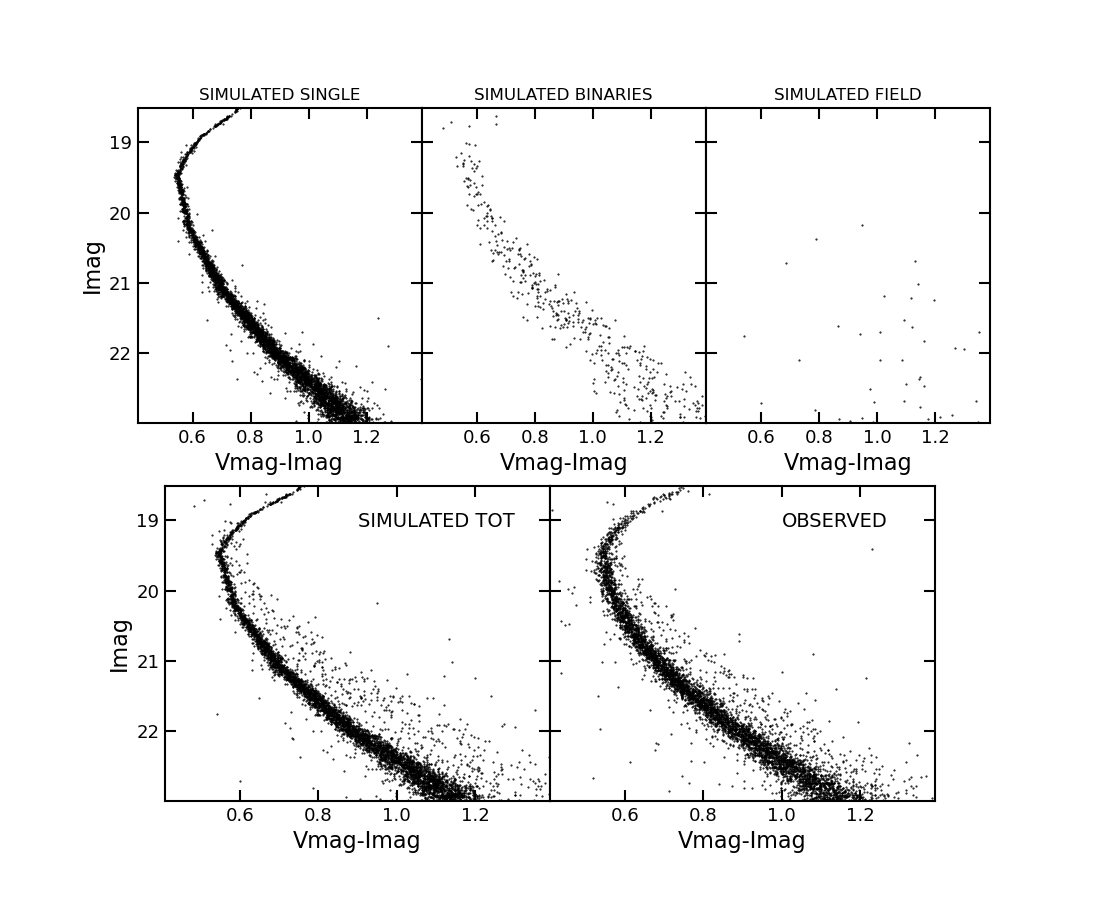}
\caption{Example of output of the total binary fraction measurement procedure applied to the entire ACS dataset of NGC 5053. The top panels show the CMDs corresponding to the simulated single main-sequence stars, binaries, and field stars. The lower panels show the comparison between the resulting synthetic CMD (left-hand panel) and the observed one (right-hand panel).}
\label{fig:ftotexample}
\end{figure}

\subsection{Results}
\label{sec:results}

\begin{figure*}[h] 
\centering
\includegraphics[scale=0.34]{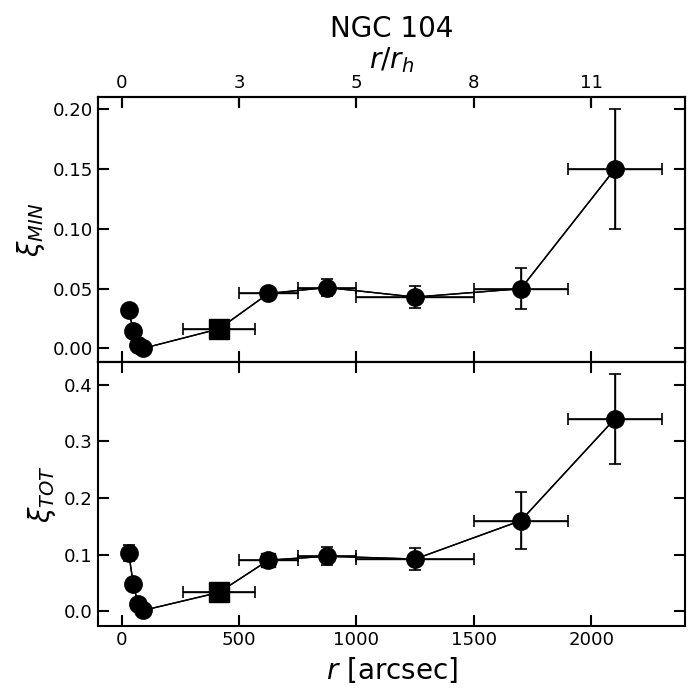}
\includegraphics[scale=0.34]{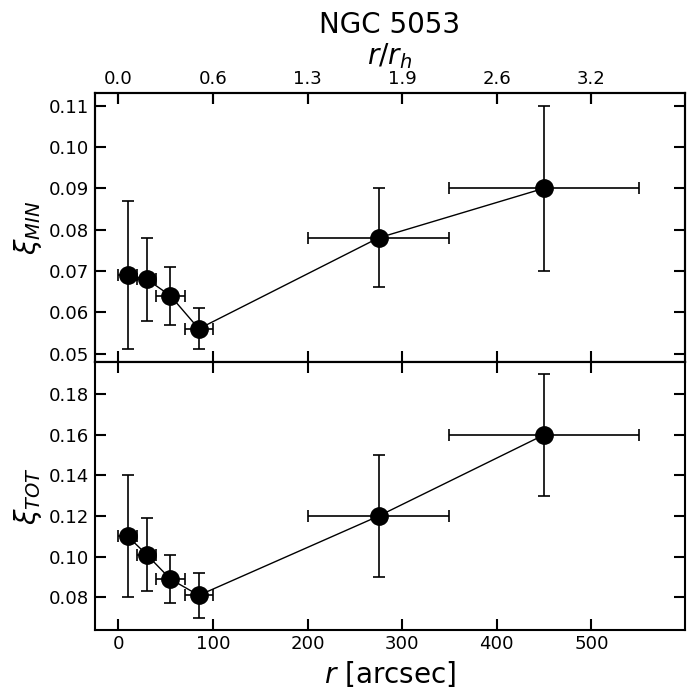}
\includegraphics[scale=0.34]{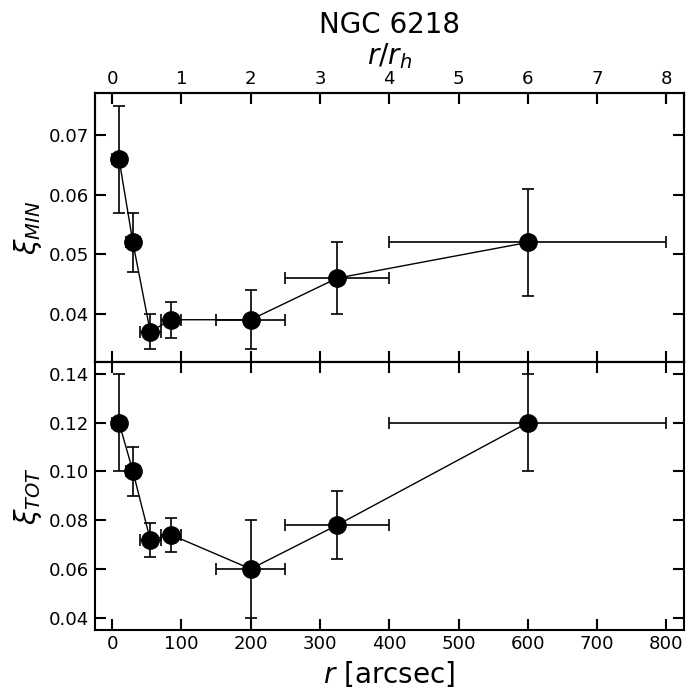}
\includegraphics[scale=0.34]{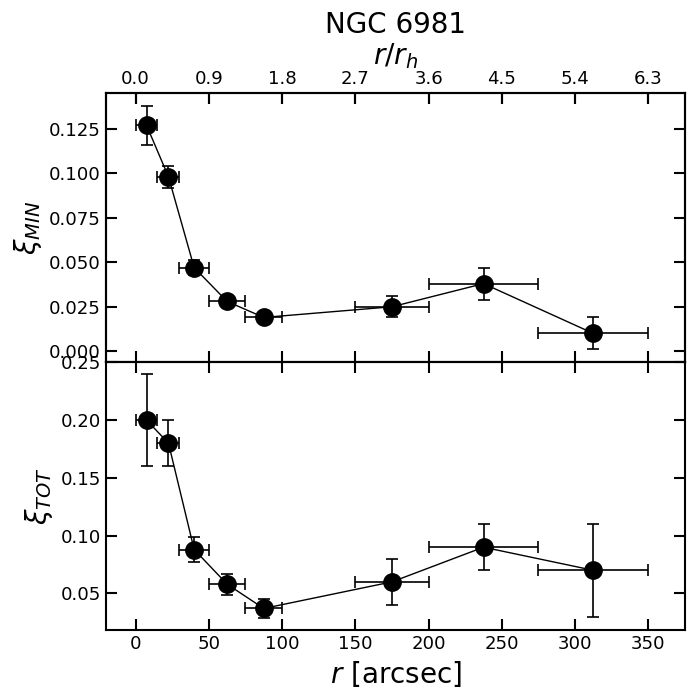}
\includegraphics[scale=0.34]{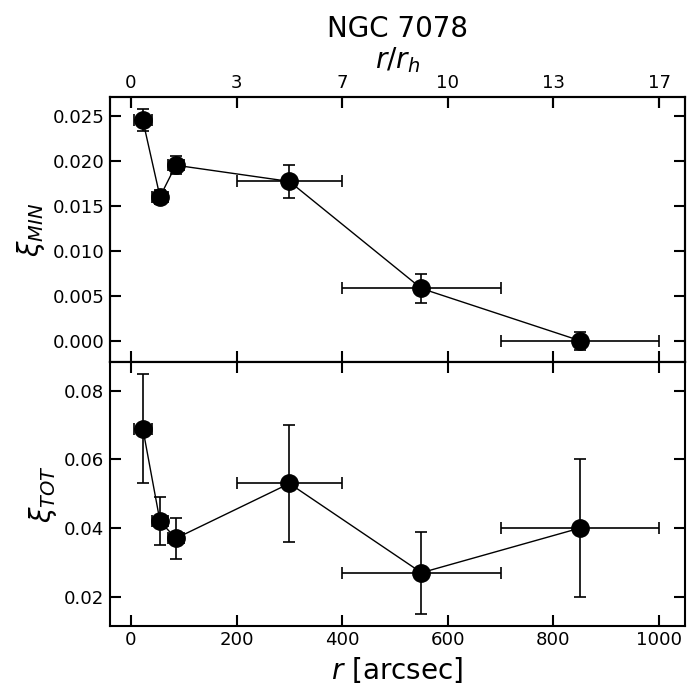}
\includegraphics[scale=0.34]{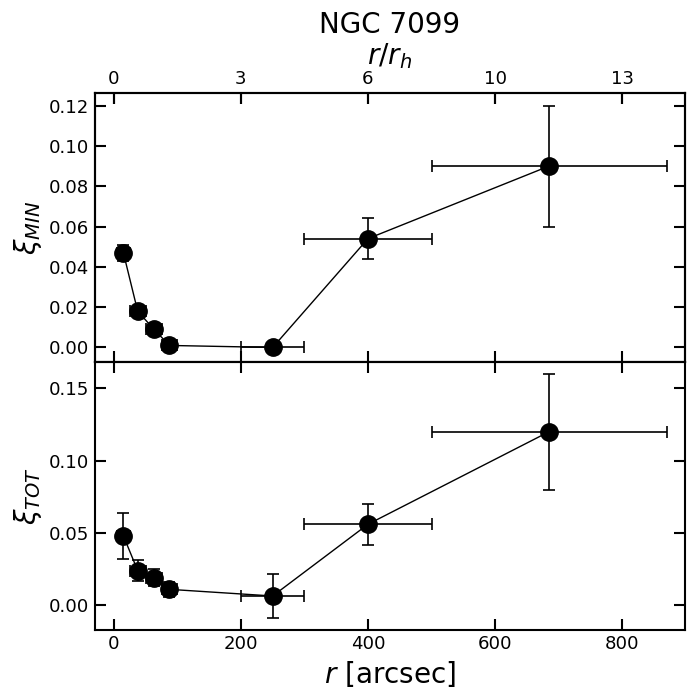}
\caption{Minimum and total binary fractions (top and bottom panel, respectively) as a function of the clustercentric distance for the investigated clusters. The square point in the case of NGC~104 is obtained through the ACS dataset from \citet{kalirai12}.}
\label{fig:fbin}
\end{figure*}

The minimum and total binary fractions for each cluster were measured in the same radial bins, extending nearly to the tidal radius. We limited the analysis to regions where the cluster main sequence is still visible in the CMDs; in some cases, no cluster stars are detected close to the tidal radius. This is not unexpected, since tidal radius estimates are highly uncertain and do not provide a reliable measure of the actual cluster extent. For NGC~104, the binary fraction in the \citet{kalirai12} ACS field was measured 2.5–5.5 mag below the turn-off, due to the lack of brighter stars in the available artificial-star catalog. This should not bias our results, as the binary fraction in old GCs shows little dependence on stellar mass \citep[e.g.,][]{dalessandro11,milone12}.

{The adopted radial binning differs between the central HST datasets and the outer wide-field observations. In the central regions, where the stellar number densities are very high, we adopted an approximately uniform radial sampling. In the outer regions, the radial bins were chosen as a compromise between maximizing the radial resolution and ensuring a robust estimate of the field contamination. In particular, the area of each bin was required to remain significantly smaller than that of the corresponding reference field used for the statistical subtraction of contaminating objects.}

The results are presented in Figure~\ref{fig:fbin}. The results also show that, for all clusters, the binary fraction exhibits a significant radial variation. Assuming a flat binary distribution at cluster formation {(i.e., same binary fraction at any clustercentric distance)}, this suggests that dynamical effects have shaped the spatial distribution of binary stars. 
The most striking feature in all clusters but NGC7078 is the presence of a bimodal distribution. The binary fraction increases toward the cluster center, confirming that mass segregation has already affected the inner regions. Central values range from $\xi_{MIN}=0.07$–$0.14$ and $\xi_{TOT}=0.12$–$0.30$. This central peak is followed by a steep decline, reaching a minimum of $\xi_{MIN}=0.00$–$0.055$ and $\xi_{TOT}=0.00$–$0.08$, before rising again in the outer regions, where values are comparable to, or even higher than, those in the cluster center. For example, in NGC104 the total binary fraction reaches a minimum compatible with 0 and then increases in the outer bins to values around 0.1–0.3, comparable to the central fraction. Interestingly, consistent with our findings, \citet{cordoni25} recently reported independent evidence for an increase in the binary fraction from the center to the outskirts of NGC104, based on Rubin Observatory Data Preview 1. A similar behavior is also observed in NGC6218. In NGC5053 and NGC7099, the total binary fraction in the outer bins reaches values $\sim50$–$150\%$ higher than in the central bin. By contrast, in NGC6981 the outer increase is more moderate: after reaching a minimum of $\sim0.04$, the total fraction rises to $\sim0.1$.

This is the first time such bimodality in the distribution of binaries has been firmly detected. {Previous studies revealed only a possible hints of this feature (see, e.g., the case of NGC~5466 in \citealt{beccari13} and Figures~34-35 in \citealt{milone12})}. The only exception is represented by NGC~7078, where the binary fraction decreases monotonically from a pronounced central peak to an approximately flat profile in the outer regions. 

 {A direct comparison of our results with previous studies requires some caution because our analysis is based on a much finer radial binning aimed at resolving local variations of the binary fraction, whereas previous works generally measured average binary fractions over much larger cluster regions. As a consequence, the strong radial gradients revealed by our analysis are largely washed out in previous measurements. Nevertheless, when our binary fractions are averaged over radial ranges comparable to those adopted by \citet{milone12}, we find overall consistency with their results. It is also worth noting that this comparison is only meaningful for the minimum binary fractions, since that study measured binaries with mass ratios $q>0.5$, whereas our analysis generally reaches slightly lower minimum detectable mass ratios. Moreover, the total binary fractions are not directly comparable because they were obtained by extrapolating the observed $q>0.5$ population assuming a flat mass-ratio distribution, while our values are derived through Monte Carlo simulations that explicitly model the CMD distribution of binaries.}
 For NGC 104, \citet{mullerhorn25} derived a spectroscopic binary fraction of $0.024\pm0.010$ from multi-epoch MUSE observations. Restricting our analysis to the overlapping radial range between their and our observations, we obtain total binary binary fraction of $0.035\pm0.006$, broadly consistent with their result within the uncertainties.} 


\section{Discussion}

{From the observational side, in order to explore the link between the evolution of the bimodal distribution of the binary fraction and the clusters' dynamical ages, we analyzed the position of the minimum in each cluster as a function of the clusters' dynamical ages (defined as the ratio of the cluster's chronological age to its relaxation timescale). To this end, we defined the position of the minimum of the binary distribution as the radial bin where the lowest total binary fraction is observed, for the five clusters showing a bimodal profile. Figure~\ref{fig:min} shows the positions of the minima in units of core radii (top panels) and in units of half-light radii (bottom panels) as a function of
the dynamical ages calculated using either the central relaxation timescale (left-hand panels) or the half-mass relaxation timescale (middle panels). 
We also included the minimum tentatively identified in NGC 5466 by \citet{beccari13}. {The statistical significance of these trends is quantified through the Spearman rank correlation coefficients and their associated p-values, reported in Figure~\ref{fig:min}. Our data show a clear correlation between the position of the minima and the cluster’s dynamical ages, as confirmed by both the Spearman coefficients and the corresponding p-values. The correlations involving $r_{\rm min}/r_h$ appear slightly weaker than those involving $r_{\rm min}/r_c$, suggesting that the core radius may provide a more robust indicator for tracing the dynamical effects acting on binaries.
The correlations demonstrate that dynamically younger clusters are characterized by minima closer to the center than more evolved systems. Moreover, they are consistent with the trend between the position of the minimum and the cluster’s dynamical age found in  the simulations presented in paper II and supports the interpretation of this trend as the result of two-body relaxation  on binaries (binary disruption and segregation of the surviving binaries; see paper II for further details).}
In the bottom panels we report between parenthesis the Spearman coefficient excluding NGC~104. In fact, when the position of the minima are normalized to the half-light radii, NGC~104 appears as a clear outlier. The origin of this discrepancy is in the large half-light of this cluster with respect to its core radius. The  physical reasons of this is beyond the scope of the paper. {Moreover, it is worth stressing that the core radius of NGC~7099 should be considered with caution, as the cluster is core-collapsed and therefore has a density profile for which the definition of a core radius is not trivial, and the value is likely underestimated.}

We also compared the position of the minimum with the $A^+_{rh}$ parameter \citep[][and references therein]{alessandrini16,ferraro23}, which quantifies the degree of central segregation of a given stellar population (such as blue straggler stars) with respect to a reference population of lighter stars within the cluster half-mass radius. This parameter is a fully empirical measurement, defined as the area enclosed between the cumulative radial distribution of the investigated population and that of the reference sample, and it is expected to increase with the dynamical ageing of the host system, as a result of the progressive sedimentation driven by dynamical friction (see Figure~7 in \citealt{ferraro23}). The $A^+_{rh}$ values for the clusters analyzed here are taken from \citet{lanzoni16,ferraro18}, except for NGC~5053 and NGC~6218, for which we provide the first measurements in this work. These were derived following the prescriptions described in \citet{ferraro18,ferraro23}, combining HST catalogs from \citet{nardiello18} with wide-field multi-band ground-based catalogs from \citet{stetson19}. Briefly, blue straggler stars were selected within the half-light radius and brighter than the evolutionary track of stars $0.2M_{\odot}$ more massive than the turn-off. As a reference population, we adopted main-sequence turn-off stars, and we constructed the cumulative radial distributions of both samples. The area between the two distributions defines $A^+_{rh}$, with uncertainties estimated through bootstrapping. Using this method, we find $A^+_{rh}=0.06\pm0.06$ for NGC~5053 and $0.16\pm0.06$ for NGC~6218. The right-hand panels of Figure~\ref{fig:min} show the position of the minima as a function of the $A^+_{rh}$ parameters. Once again, a strong correlation emerges,  indicating that the farther the binary fraction minimum is displaced from the center, the larger the $A^+_{rh}$ value, meaning that these systems have been more strongly affected by long-term mass segregation effects.

\begin{figure*}[h] 
\centering
\includegraphics[scale=0.5]{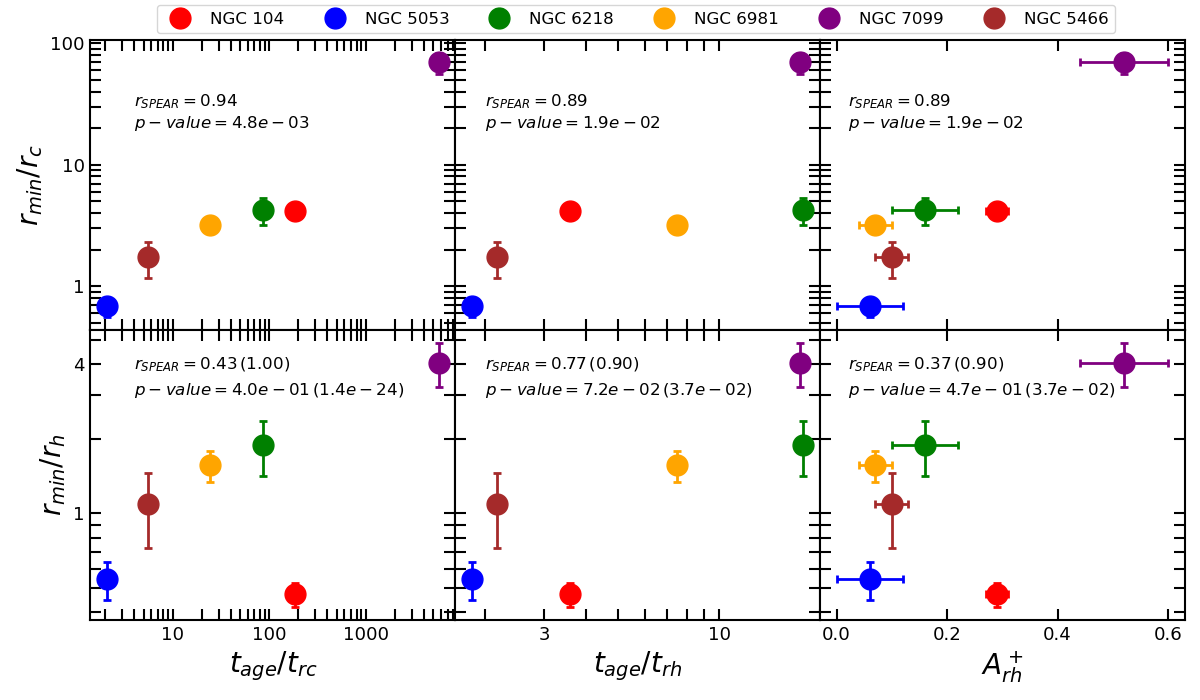}
\caption{
Position of the minimum of the radial distribution of the total binary fraction as a function of the cluster dynamical age.
Top panels: minima expressed in units of core radius.
Bottom panels: minima expressed in units of half-light radius.
The dynamical age is parametrized as the ratio between the cluster chronological age and the central relaxation time (left panels), the half-mass relaxation time (middle panels), and through the empirical parameter \(A^+_{rh}\) derived from the radial distribution of blue straggler stars (right panels).
Different colors correspond to different GCs, as indicated in the top legend.
The Spearman rank correlation coefficients are reported in each panel, with values in parentheses referring to the correlation obtained excluding NGC~104.
}
\label{fig:min}
\end{figure*}

{
Although these results support a general picture in which local variations in the binary fraction are linked to the degree of internal dynamical evolution, some caveats are required. 
First, while the monotonic distribution observed in NGC~7078 is in principle consistent with its advanced dynamical stage as a core-collapsed cluster, NGC~7099 instead shows a striking bimodality despite being a core-collapsed system as well. 
For non-core-collapsed clusters, it is interesting to note the case of dynamically young cluster NGC 6101, for which \citet{dalessandro15} found a flat binary fraction distribution, not observed in any of the clusters in our sample. Altogether, these findings suggest that additional dynamical ingredients may, in some cases, be responsible for shaping the spatial distribution of binaries within GCs. 
For example, \citet{alessandrini16b} and \citet{peuten16} discussed the possible role of dark stellar remnants in delaying the mass segregation of visible stars and binaries in clusters such as NGC~6101, while \citet{gill08} showed that mass segregation can also be slowed by the presence of an intermediate-mass black hole (see also \citealt{webb17} for a study of mass segregation and the stellar initial mass function in NGC~6101 and other clusters).
}

{
The cases of NGC~7078 and NGC~7099 deserve particular attention.
Both clusters are classified as extremely dynamically evolved systems, having already undergone core collapse \citep{ferraro09,beccari19}.
Based on their dynamical state and on the comparison with blue straggler star distributions, one would expect a unimodal and strongly segregated binary profile in both systems; however, this behaviour is observed only in NGC~7078.
The origin of this discrepancy is puzzling, especially considering that both the $t/t_{rc}$ and $t/t_{rh}$ values of NGC~7099 are significantly larger than those of NGC~7078.
One possible explanation is that the combination of high central densities and strong mass segregation in such an evolved system may create an efficient “scattering machine” that ejects binaries from the core to wider orbits, thereby replenishing the cluster outskirts.
Indeed, in one of the simulations presented in Paper~II we show that, after a cluster undergoes deep core collapse, the scattering of binaries from the inner to the outer regions can regenerate a rising branch in the outer radial profile of the binary fraction.
This process, however, requires additional time after the core collapse ($\sim2$ Gyrs in the simulation presented in paper II), which could explain why a bimodal profile is observed only in the dynamically older of the two post-core-collapse clusters in our sample.
}
Interestingly, it is worth mentioning that a similar mechanism may be at work in the core-collapsed GC NGC~6752, which shares comparable structural parameters with NGC~7099. In that case, the anomalous positions of two binary millisecond pulsars \citep{corongiu24} can only be explained by their formation in the cluster core followed by subsequent dynamical ejection \citep{phinney91,colpi02}.

\section{Summary and Conclusions} \label{sec:conc}

In this work, we conducted the first homogeneous and systematic study of the binary fraction across the entire radial extent of six representative Galactic GCs spanning a broad range of dynamical ages. Our sample includes clusters from dynamically young systems like NGC~5053 to those in an advanced stage of evolution, such as the two core-collapse systems NGC~7078 and NGC~7099. This approach has allowed us to coherently investigate how internal dynamical processes shape the distribution of binary systems in different environments. {We find that that the binary fraction varies significantly with the clustercentric distance. The most intriguing feature observed in the binary radial distribution is the presence of a striking bimodal profile which we find in 5 out of the 6 investigated clusters. In these systems, the binary fraction is characterized by a central peak, a minimum at intermediate radii, and a rising branch in the outer regions.}

{Our data show a significant
correlation between the position of the minima and
the cluster (core and half-mass) relaxation timescales indicating that the cluster's internal dynamical processes play a major role in shaping the binary fraction's radial profile; a strong
correlation has also been observed with the $A^+_{rh}$, an empirical measurement of the cluster dynamical age based on the degree of mass segregation of blue straggler stars.
In paper II we present the results of a set of Monte Carlo
simulations following the evolution of GCs and their binary
star population. Our simulations show that binary bimodal
spatial distributions consistent with those found in our observations can emerge as a result of the combined effect
of binary disruption and segregation of the surviving binaries in cluster hosting multiple stellar populations. Moreover,
single–binary interactions in the extremely dense cores of
post–core-collapse clusters may also play a key role in shaping the radial profile of the binary fraction in clusters which have reached the post-core-collapse phase.}

Looking ahead, we aim to expand the observational analysis to a larger sample of clusters to provide further observational constraints on the origin of the bimodality and its link with the clusters' evolutionary history, its present-day properties, and the implications for our understanding of exotic stellar populations assicated with the binary stars like blue straggler stars, cataclysmic variables, and millisecond pulsars. The results presented here represent a crucial first step toward a more complete picture of binary star evolution and cluster dynamics, demonstrating the power of binary systems as probes of the complex interplay between stellar evolution and dynamical processes in dense stellar environments.


\begin{acknowledgements}
We thank the referee for the careful reading of the manuscript and for the many constructive comments and suggestions which improved the manuscript quality. This work is part of the project Cosmic-Lab (Globular
Clusters as Cosmic Laboratories) at the Physics and Astronomy Department “A. Righi” of the Bologna University (\url{http://www.cosmic-lab.eu/Cosmic-Lab/Home.html}) and it has been supported  by the project GENESIS - Searching for the primordial structures of the Universe in the heart of the Galaxy (Advanced Grant FIS-2024-02056, PI:Ferraro), funded by the Italian MUR through the Fondo Italiano per la Scienza call.
ED acknowledges financial support from the INAF Data analysis Research Grant (PI E. Dalessandro) of the “Bando Astrofisica Fondamentale 2024''. 

\end{acknowledgements}

\bibliographystyle{aa}
\bibliography{biblio}{}

\end{document}